\documentclass[12pt,a4paper]{article}

\usepackage{amssymb,amsmath,eucal}
\usepackage[dvips]{lscape,graphicx}
\usepackage{cite}

\renewcommand{\thefootnote}{\arabic{footnote}}

\newcommand{\bc}{\begin{center}}
\newcommand{\ec}{\end{center}}
\newcommand{\bd}{\begin{displaymath}}
\newcommand{\ed}{\end{displaymath}}
\newcommand{\be}{\begin{equation}}
\newcommand{\ee}{\end{equation}}
\newcommand{\ba}{\begin{array}}
\newcommand{\ea}{\end{array}}
\newcommand{\bt}{\begin{tabular}}
\newcommand{\et}{\end{tabular}}
\newcommand{\un}{\underline}
\newcommand{\ov}{\overline}
\newcommand{\ds}{\displaystyle}

\newcommand{\ct}{\cite}

\newcommand{\lb}{\label}
\newcommand{\bp}{\begin{picture}}
\newcommand{\ep}{\end{picture}}
\newcommand{\bfi}{\begin{figure}}
\begin{document}

\title{\huge\bf {Monopoles and Family Replicated Unification}}

\author{\\{\bf L.V.Laperashvili\footnotemark[1]}\\[0.3cm]
{\bf H.B.Nielsen{\renewcommand{\thefootnote}{\arabic{footnote})}\footnotemark[1]}\;\,
\footnotemark[7]}\\[0.2cm]
{\bf D.A.Ryzhikh{\renewcommand{\thefootnote}{\arabic{footnote})}\footnotemark[2]}\;\;\;%
\footnotemark[1]\:\,\footnotemark[1]\:\,\footnotemark[1]}\\[1cm]
\it Institute of Theoretical and Experimental Physics,\\
\it Moscow, Russia}

\date{}

{\renewcommand{\thefootnote}{\arabic{footnote})}
\addtocounter{footnote}{+1}
\footnotetext{Niels Bohr Institute, Copenhagen, Denmark.}}

{\renewcommand{\thefootnote}{\arabic{footnote})}
\addtocounter{footnote}{+1}
\footnotetext{ Institute of Theoretical and Experimental Physics, Moscow, Russia.}}

{\renewcommand{\thefootnote}{\fnsymbol{footnote}}
\addtocounter{footnote}{-1}
\footnotetext{{\bf E-mail}: laper@heron.itep.ru}}

{\renewcommand{\thefootnote}{\fnsymbol{footnote}}
\addtocounter{footnote}{+6}
\footnotetext{{\bf E-mail}: hbech@alf.nbi.dk}}

{\renewcommand{\thefootnote}{\fnsymbol{footnote}$*$}
\footnotetext{{\bf E-mail}: ryzhikh@heron.itep.ru}}

\maketitle

\thispagestyle{empty}

\newpage
\begin{abstract}
The present theory is based on the assumption that at the
very small (Planck scale) distances our space-time is
discrete, and this discreteness influences on the Planck
scale physics. Considering our (3+1)-dimensional space-time
as a regular hypercubic lattice with a parameter
$a=\lambda_\text{P}$, where $\lambda_\text{P}$ is the Planck length, we
have investigated a role of lattice artifact monopoles
which is essential near the Planck scale if the Family
replicated gauge group model (FRGGM) is an extension of the
Standard Model at high energies. It was shown that
monopoles have $N$ times smaller magnetic charge in FRGGM
than in SM ($N$ is the number of families in FRGGM).
These monopoles can give an additional contribution to
beta-functions of the renormalisation group equations for
the running fine structure constants $\alpha_\text{i}(\mu)$ (i=1,2,3
correspond to the U(1), SU(2), and SU(3) gauge groups of the Standard
Model). We have used the Dirac relation for renormalised
electric and magnetic charges. Also we have estimated the
enlargement of a number of fermions in FRGGM leading to
the suppression of the asymptotic freedom in the
non-Abelian theory. Different role of monopoles in the
vicinity of the Planck scale gives rise or to AntiGUT, or
to the new possibility of unification of gauge interactions
(including gravity) at the scale $\mu_\text{GUT}\approx
10^{18.4}$ GeV. We discussed the possibility of the
[SU(5)]$^3$ SUSY or [SO(10)]$^3$ SUSY unifications.
\end{abstract}

\thispagestyle{empty}

\newpage

\pagenumbering{arabic}

\section{Introduction}

Trying to look insight the Nature and considering the physical processes
at very small distances, physicists have made attempts to explain the
well--known laws of low--energy physics as a consequence of the more
fundamental laws of Nature. The contemporary physics of the electroweak
and strong interactions is described by the Standard Model (SM) which unifies
the Glashow-Salam-Weinberg electroweak theory with QCD -- theory of strong
interactions.

The gauge group of symmetry in the SM is :
\begin{equation}
SMG = SU(3)_\text{c}\times SU(2)_\text{L}\times U(1)_\text{Y},
\lb{1}
\end{equation}
which describes the present elementary particle physics up to the scale
$\approx 100$ GeV.

Recently it was shown in a number of papers~\cite{o1} that the family replicated gauge groups of type :
\be
SU(n)^N\times SU(m)^N
\lb{2}
\ee
play an essential role in construction of renormalizable, asymptotically
free, four dimensional gauge theories that dynamically generate a fifth
dimension (or fifth and sixth ones). This theory leads to the natural
electroweak symmetry breaking, relying neither on supersymmetry nor on
strong dynamics at the TeV scale. The new TeV physics is perturbative, and
radiative corrections to the Higgs mass are finite. The Higgs scalar is an
extended object -- pseudo-Nambu-Goldstone boson -- and a novel
Higgs potential emerges naturally requiring a second light SU(2)
doublet scalar.

We see that the family replicated gauge groups provide a new way
to stabilize the Higgs mass in the Standard Model.

\section{Family Replicated Gauge Group}

The extension of SM with the Family Replicated Gauge Group (FRGG):
\begin{equation}
G = (SMG)^N = [SU(3)_\text{c}]^N\times [SU(2)_\text{L}]^N\times [U(1)_\text{Y}]^N
\lb{3}
\end{equation}
was first suggested by C.D.Froggatt and
H.B.Nielsen~\cite{I1}.

In Eq.(\ref{3}) $N$ designates the number of quark and lepton families. If
$N=3$
(as experiment confirms), then the fundamental gauge group G is:
\begin{equation}
G = (SMG)^3 = SMG_\text{1st\;fam.}\times SMG_\text{2nd\;fam.}\times SMG_\text{3rd\;fam.}.
\lb{4}
\end{equation}
The generalized fundamental group:
\begin{equation}
G_\text{f} = (SMG)^3\times U(1)_\text{f}
\lb{5}
\end{equation}
was suggested by the fitting of fermion masses of the
SM~\cite{I2}.


Recently a new generalization of FRGG-model was suggested in
papers~\cite{I3}, in which the fundamental group:
\begin{equation}
\begin{array}{c}
G_{\mbox{ext}} = (SMG\times U(1)_\text{B-L})^3\\
\equiv [SU(3)_\text{c}]^3\times [SU(2)_\text{L}]^3\times [U(1)_\text{Y}]^3\times [U(1)_\text{(B-L)}]^3
\end{array}
\lb{6}
\end{equation}
takes into account the see-saw mechanism with right-handed
neutrinos, describes all modern neutrino experiments, and gives
the reasonable fitting of the SM fermion masses and mixing angles. The group
$G=G_\text{ext}$ contains: $3\times 8 = 24$ gluons, $3\times
3 = 9$ W-bosons and $3\times 1 + 3\times 1 = 6$ Abelian gauge
bosons.

The model is renormalisable: has no anomalies, neither gauge nor
mixed.

The gauge group $G_{\mbox{ext}}$ undergoes the spontaneous breakdown
(at some orders of magnitude below the Planck scale)
by 7 different Higgs fields to the gauge group
which is the diagonal subgroup of $G_{\mbox{ext}}$.
Therefore, 7 Higgs fields break FRGG-model to the SM.  The field
$\phi_\text{WS}$ corresponds to the Weinberg-Salam theory.
Its VEV is known: $<\phi_\text{WS}>=246$ GeV, so that
we have only 6 free parameters -- six VEVs --
to fit the experiment in the framework of this model.

Froggatt, Nielsen and Takanishi~\cite{I3}
have used them with aim to find the best fit to conventional
experimental data for all fermion masses and mixing angles in the SM, also
to explain the experiments in the neutrino oscillations. The typical fit was encouraging in the crude approximation.
Also the neutrino masses were predicted.

Finally, we conclude that, in general, the theory with
FRGG-symmetry is successful in describing of the SM
experiment.

\section{Lattice-like Structure of our Space-Time}

Having an interest in the fundamental laws of physics, we can consider
the two possibilities:
\begin{itemize}
\item[1.] At the very small (Planck length) distances
{\un{our space-time is continuous}}
and there exists the fundamental theory with a very high symmetry.
\item[2.] At the very small distances {\un{our space-time is
discrete}}, and this discreteness influences on the Planck scale physics.
\end{itemize}
The item 2 is an initial (basic) point of view of the present theory, but not
an approximation. It is a base of the theory of physical processes proceeding
at small distances of order of the Planck scale $\lambda_\text{P}=M_\text{Pl}^{-1}$:
\begin{equation}
M_\text{Pl}=1.22\cdot 10^{19}\,{\mbox{GeV}}.
\lb{7}
\end{equation}

In the simplest case we can imagine our (3+1) space--time as a
regular hypercubic lattice with a parameter $a=\lambda_\text{P}$.
Then the lattice artifact monopoles can play an essential role near the
Planck scale. But, of course, it is necessary to comment that we
do not know (at least on the level of our today knowledge) what
lattice-like structure (random lattice, or foam, or string
lattice, etc.) plays role in the description of physical processes
at very small distances~{\cite{I4}}.

The aim of the present paper is also to show that monopoles cannot be seen
in the usual SM up to the Planck scale, because they have a huge magnetic
charge and are completely confined. Supersymmetry does not help to see monopoles.

We suggest to consider a possibility of the existence of monopoles
in our World, extending the Standard Model Group (SMG) to the Family Replicated
Gauge Group (SMG)$^N$.

\section {Renormalisation Group Equations for Electric and Magnetic
Fine Structure Constants}

J.Schwinger~\cite{I5} was first who investigated the problem of renormalisation
of the magnetic charge in Quantum ElectroMagnetoDynamics (QEMD), i.e.
in the Abelian quantum field theory of electrically and magnetically
charged particles (with charges $e$ and $g$, respectively).

Considering the "bare" charges $e_\text{0}$ and $g_\text{0}$ and renormalised (effective)
charges $e$ and $g$, Schwinger obtained:
\begin{equation}
e/g = e_\text{0}/g_\text{0},
\lb{8}
\end{equation}
what means the absence of the Dirac relation for the renormalised
electric and magnetic charges.

But there exists another solution of this
problem~[\citen{I6}--\citen{I8}] which gives:
\begin{equation}
eg = e_\text{0}g_\text{0} = 2\pi n,\quad (n\in Z),
\lb{9}
\end{equation}
i.e. the existence of the Dirac relation (charge quantization condition) for
both, bare and renormalised electric and magnetic charges.
Here we have $n=1$ for the minimal (elementary) charges.

These two cases lead to the two possibilities for the renormalisation group
equations (RGEs) describing the evolution of electric and magnetic fine
structure constants:
\begin{equation}
\alpha = \frac{e^2}{4\pi}\quad{\mbox{and}}\quad
\widetilde \alpha = \frac{g^2}{4\pi},
\lb{10}
\end{equation}
which obey the following RGEs containing the electric and magnetic
beta-functions:
\begin{equation}
\frac {d(\log \alpha(\mu))}{dt} = \pm \frac {d(\log \widetilde \alpha(\mu))}{dt}
= \beta^{(e)}(\alpha) \pm \beta^{(m)}(\widetilde \alpha).
\lb{11}
\end{equation}
In Eq.(\ref{11}) we have:
\begin{equation}
t = \log(\frac{\mu^2}{\mu_R^2}),
\lb{12}
\end{equation}
where $\mu$ is the energy scale and $\mu_\text{R}$ is the renormalisation point.

The second possibility (with minuses) in Eq.(\ref{11})
corresponds to the validity of the Dirac relation for the
renormalised charges.  We believe only in this case considered by authors
in paper~\cite{I8} where we have used the Zwanziger formalism of QEMD~\cite{I9}.

In the present paper, excluding the Schwinger's
renormalisation condition~(\ref{8}), we assume only the Dirac relation
for running $\alpha$ and $\widetilde \alpha$:
\be
\alpha \widetilde \alpha = \frac{1}{4}.
\lb{13}         
\ee
It is necessary to comment that RGEs (\ref{11}) are valid only for
$\mu > \mu_\text{threshold} = m_\text{mon}$, where $m_\text{mon}$ is the monopole mass.

If monopole charges, together with electric ones, are sufficiently small,
then $\beta$-functions can be considered perturbatively:
\begin{equation}
\beta(\alpha) = \beta_\text{2} (\alpha /4\pi) + \beta_\text{4} {(\alpha/4\pi)}^2 + ...
\lb{18x}
\end{equation}
and
\begin{equation}
\beta(\widetilde \alpha) = \beta_\text{2} (\widetilde \alpha /4\pi) +
\beta_4 {(\widetilde \alpha/4\pi)}^2 + ...
\lb{19x}
\end{equation}
with (see~\cite{I8} and references there)
\begin{equation}
\beta_2 = \frac 13  \quad{\mbox{and}}\quad \beta_\text{4} =1 \quad
-\quad {\mbox {for scalar particles}},
\lb{20x}
\end{equation}
and
\begin{equation}
\beta_2 = \frac 43  \quad{\mbox{and}}\quad \beta_\text{4} \approx 4
\quad-\quad {\mbox {for fermions}}.
\lb{21x}
\end{equation}
For scalar electric and magnetic charges we have:
\begin{equation}
\frac {d(\log \alpha(\mu))}{dt} = - \frac {d(\log
\widetilde \alpha(\mu))}{dt} = \beta_2 \frac{\alpha - \widetilde \alpha}{4\pi}(1 + 3
\frac{\alpha + \widetilde \alpha}{4\pi} + ...)
\lb{17x}
\end{equation}
with $\beta_\text{2} = 1/3$, and approximately the same result is valid
for fermionic particles with $\beta_\text{2} = 4/3$.
Eq.(\ref{17x}) shows that there exists a region when both fine structure
constants are perturbative. Approximately this region is given by the
following inequalities:
\begin{equation}
0.2 \stackrel{<}{\sim }(\alpha, \widetilde \alpha)
\stackrel{<}{\sim }1.
\lb{22x}
\end{equation}
Using the Dirac relation (\ref{13}), we see from Eq.(\ref{17x})
that in the region (\ref{22x}) the two-loop contribution is not larger
than 30\% of the one-loop contribution, and the perturbation theory
can be realized in this case.

It is necessary to comment that the region (\ref{22x}) almost coincides
with the region of phase transition couplings obtained in the lattice
compact QED~\cite{I10a}.

\section{Evolution of Running Fine Structure Constants}

The usual definition of the SM coupling constants is given in
{\it the Modified minimal subtraction scheme}($\ov{\text{MS}}$):
\begin{equation}
\alpha_\text{1} = \frac{5}{3}\alpha_\text{Y},\quad
\alpha_\text{Y} = \frac{\alpha}{\cos^2\theta_{\ov{\text{MS}}}},\quad
\alpha_\text{2} = \frac{\alpha}{\sin^2\theta_{\ov{\text{MS}}}},\quad
\alpha_\text{3} \equiv \alpha_\text{s} = \frac {g^2_\text{s}}{4\pi},
\lb{81y}
\end{equation}
where $\alpha$ and $\alpha_\text{s}$ are the electromagnetic and SU(3)
fine structure constants respectively, Y is the hypercharge, and
$\theta_{\ov{\text{MS}}}$ is the Weinberg weak angle in $\ov{\text{MS}}$ scheme.
Using RGEs with experimentally established parameters,
it is possible to extrapolate the experimental
values of three inverse running constants $\alpha_\text{i}^{-1}(\mu)$
(here i=1,2,3 correspond to U(1), SU(2) and SU(3) groups of SM)
from the Electroweak scale to the Planck scale.

It is well known (see for example~\cite{I10}) that (in the absence of monopoles) the one--loop
approximation RGEs can be described by the following expressions:
\begin{equation}
\alpha_\text{i}^{-1}(\mu) =
\alpha_\text{i}^{-1}(\mu_\text{R}) + \frac{b_\text{
i}}{4\pi}t,
\lb{4z}
\end{equation}
where slopes $b_\text{i}$ are given by the following values:
\begin{equation}
\begin{array}{c}
b_\text{i} = (b_\text{1}, b_\text{2}, b_\text{3}) =\\
( - \frac{4}{3}N_\text{gen} -\frac{1}{10}N_\text{S}, \quad
\frac{22}{3}N_\text{V} - \frac{4}{3}N_\text{gen} -\frac{1}{6}N_\text{S}, \quad
11 N_\text{V} - \frac{4}{3}N_\text{gen} ).
\end{array}
\lb{5z}
\end{equation}
The integers $N_\text{gen},\,N_\text{S},\,N_\text{V}$ are respectively the numbers
of generations, Higgs bosons and different vector gauge fields.

In SM we have:
\begin{equation}
N_\text{gen} = 3, \quad N_\text{S} = N_\text{V} =1,
\lb{6z}
\end{equation}
and the corresponding slopes (\ref{5z}) describe the evolutions of
$\alpha_\text{i}^{-1}(\mu)$.

The precision of the LEP data allows to make the extrapolation of RGEs
with small errors up to the Planck scale unless the
new physics pops, of course. Assuming that these
RGEs for $\alpha_\text{i}^{-1}(\mu)$ contain only the contributions of the SM
particles up to $\mu=\mu_\text{Pl}\equiv M_\text{Pl}$ and doing the extrapolation with one Higgs
doublet under the assumption of a "desert" and absence of monopoles,
we have the following result obtained in~\cite{I11}:
\begin{equation}
\alpha_\text{1}^{-1}(\mu_\text{Pl})\approx 33.3; \quad
\alpha_\text{2}^{-1}(\mu_\text{Pl})\approx 49.5; \quad
\alpha_\text{3}^{-1}(\mu_\text{Pl})\approx 54.0.
\lb{82y}
\end{equation}
The extrapolation of $\alpha_\text{1,2,3}^{-1}(\mu)$ up to the point
$\mu=\mu_\text{Pl}$ is shown in Fig.1 as function of the variable
$x=log_\text{10}\mu$ (GeV). In this connection, it is very attractive to consider
also the gravitational interaction.

The gravitational interaction between two particles
of equal masses M is given by the usual classical Newtonian potential:
\begin{equation}
V_\text{g} = - G \frac{M^2}{r} =
- \left(\frac{M}{M_\text{Pl}}\right)^2\frac{1}{r}
= - \frac{\alpha_\text{g}(M)}{r},
\lb{1x}
\end{equation}
which always can be imagined as a tree--level approximation of quantum
gravity.

Then the quantity:
\begin{equation}
      \alpha_\text{g} = \left(\frac{\mu}{\mu_\text{Pl}}\right)^2     \lb{2x}
\end{equation}
plays a role of the running "gravitational fine structure constant"
and the evolution of its inverse quantity also is presented
in Fig.1 together with the evolutions of $\alpha_\text{i}^{-1}(\mu)$.

\section{Dropping of the Monopole Charge in the Family Replicated Gauge Group
Model (FRGGM)}

In the simplest case, the scalar monopole beta-function in QEMD is (see~\cite{I12} and~\cite{I13}):
\be
\beta(\widetilde \alpha) = \frac{\widetilde \alpha}{12\pi } +
{(\frac{\widetilde \alpha}{4\pi })}^2 + ... = \frac{\widetilde \alpha}{12\pi }( 1 +
3\frac{\widetilde \alpha}{4\pi } + ...).
\lb{1d}
\ee
From the last equation it follows that the theory of monopoles cannot
be considered perturbatively
 at least for
\be
          \widetilde \alpha > \frac{4\pi}{3}\approx 4.       \lb{2d}
\ee
This limit is smaller for non-Abelian monopoles.

Using the Dirac relation, it is easy to estimate in the simple SM
the Planck scale value $\widetilde \alpha(\mu_\text{Pl})$
(minimal for $U(1)_\text{Y}$ gauge group):
\be
\widetilde \alpha(\mu_\text{Pl}) = \frac{5}{3}\alpha_\text{1}^{-1}(\mu_\text{Pl})/4
\approx 55.5/4 \approx 14.
\lb{3d}
\ee
This value is really very big compared with the estimate (\ref{2d}) and, of
course, with the critical coupling $\widetilde \alpha_\text{crit}\approx 1$,
corresponding to the confinement-deconfinement phase
transition in the lattice  QED~\cite{I10a}. Clearly we cannot do the perturbation approximation with such
a strong coupling $\widetilde{\alpha}$.

It is hard for such monopoles not to be confined.

There is an interesting way out of this problem if one wants to
have the existence of monopoles, namely to extend the SM gauge
group so cleverly that certain selected linear
combinations of charges get bigger electric couplings than the corresponding
SM couplings. That could make the monopoles which for these
certain charge linear combinations couple more weakly and thus
have a better chance of being allowed "to exist".

An example of such an extension of SM that can impose the
possibility of the allowance of monopoles is just Family
Replicated Gauge Group Model (FRGGM).

According to the FRGGM, at some point $\mu=\mu_\text{G} < \mu_\text{Pl}$
(or really in a couple of steps)
the fundamental group $G  \equiv  G_{\mbox{ext}} $
undergoes spontaneous breakdown to its diagonal subgroup:
\begin{equation}
G \longrightarrow G_\text{diag.subgr.} = \{g,g,g || g\in SMG\},
\lb{83y}
\end{equation}
which is identified with the usual (low-energy) group SMG.

It should be said that in the FRGG-model
each family has its own gluons, own W's, and own photons. The breaking just
makes linear combination of a certain color combination of gluons which
exists in the SM below $\mu=\mu_\text{G}$ and down to the low energies.
We can say that the phenomenological gluon is a linear
combination (with amplitude $1/\sqrt 3$ for $N=3$) for each of the
FRGG gluons of the same color combination.
Then we have the following formula connecting the fine structure constants
of non-Abelian FRGG-model and low energy surviving diagonal subgroup
$G_{\mbox{diag.subg.}}\subseteq {(SMG)}^3$:
\begin{equation}
\alpha_\text{i,\mbox{diag}}^{-1} = \alpha_\text{i,\mbox{1st\, fam.}}^ {-1} +
\alpha_\text{i,\mbox{2nd\, fam.}}^{-1} + \alpha_\text{i,\mbox{3rd\,
fam.}}^{-1}.
\lb{86yA}
\end{equation}
Here i = SU(2), SU(3), and i=3 means that we talk about the gluon
couplings.

Assuming that three FRGG couplings are equal to each other, we obtain:
\begin{equation}
\alpha_\text{i,\mbox{diag}}^{-1}\approx 3\alpha_\text{i,\mbox{one fam.}}^
{-1} \equiv 3\alpha_\text{i,G}^{-1}.
\lb{86yB}
\end{equation}
In contrast to non-Abelian theories, in which the gauge invariance
forbids the mixed (in families) terms in the Lagrangian of
FRGG-theory, the U(1)-sector of FRGG contains such mixed
terms:
\begin{equation}
\frac{1}{g^2}\sum_\text{p,q} F_{\mu\nu,\;\text{p}}F_\text{q}^{\mu\nu} =
\frac{1}{g^2_\text{11}}F_{\mu\nu,\; 1}F_{ 1}^{\mu\nu} +
\frac{1}{g^2_\text{12}}F_{\mu\nu,\; 1}F_{ 2}^{\mu\nu} +
...
+ \frac{1}{g^2_\text{23}}F_{\mu\nu,\; 2}F_\text{3}^{\mu\nu} +
\frac{1}{g^2_\text{33}}F_{\mu\nu,\; 3}F_\text{3}^{\mu\nu},
\lb{87y}
\end{equation}
where p,q = 1,2,3 are the indices of three families of the group
(SMG)$^3$. Now it is easily seen that if the different families
had specific equal electric charges, i.e. equal $\alpha_\text{pq}$,
then taking the diagonal subgroup we get~\cite{I14}:
\begin{equation}
\alpha_{\mbox{diag}}^{-1}\approx 6\alpha_{\mbox{G}}^ {-1},
\lb{87yB}
\end{equation}
which shows that we can increase electric $\alpha$ by a factor 6
replacing it by the electric $\alpha_{\mbox{one\, fam.}}\equiv
\alpha_{\text G}$.

Taking (\ref{87yB}), we can get the monopole fine structure constant $\widetilde
\alpha_{\text G}$ which is smaller by factor 6 in comparison with $\widetilde \alpha$ in
the SM.  We can estimate at the Planck scale:
\be
\widetilde \alpha_G(\mu_{Pl}) \approx 14/6 \approx 2.3 .
\lb{4d}
\ee
But it seems (see below) that in the FRGGM we have
at the Planck scale:
$$
\widetilde \alpha_{G}(\mu_{Pl}) \approx 1,
$$
and the perturbation theory works for $\beta$-function of scalar
monopoles near the Planck scale.

The conclusion: if one wants monopoles "to exist", it is necessary
to drive in the direction of a model like FRGG.

\section{The possibility of Grand Unification Near the Planck Scale}

In the AntiGUT-model by Froggatt and
Nielsen~[\citen{I1}--\citen{I3}]
the FRGG breakdown was considered at $\mu_{\text G}\sim 10^{18}$ GeV.

But the aim of this investigation is to show that
we can see quite different consequences of the extension of SM to FRGGM
if G-group undergoes the breakdown to its diagonal subgroup (i.e. SM)
not at $\mu_{\text G}\sim 10^{18}$ {GeV}, but at $\mu_{\text G}\sim 10^{14}$ or $10^{15}$
{GeV}, i.e. before the intersection of $\alpha_{\text 2}^{-1}(\mu)$ with
$\alpha_{\text 3}^{-1}(\mu)$ at $\mu\approx 10^{16}$ GeV.

Then in the region $\mu_{\text G} < \mu < \mu_{\text{Pl}}$ we have three
$SMG\times U(1)_{\text{B-L}}$ groups for three FRGG families.
In this region we have a lot of fermions, mass protected or not mass protected,
belonging to usual families or to mirror ones. In FRGGM the additional 6 Higgs bosons, with their large VEVs, are
responsible for the mass protection of a lot new fermions appearing in the
region $\mu > \mu_\text{G}$. In this region we designate the total number of fermions
$N_\text{F}$, which is different with $N$.

Also a role of artifact monopoles can be important in
the vicinity of the Planck scale. Lattice monopoles are responsible for the confinement in lattice gauge
theories what is confirmed by many numerical and theoretical investigations
(see review \cite{I15} and papers \ct{I16}).
In the compact lattice gauge theory the monopoles are not physical objects:
they are lattice artifacts driven to infinite mass in the continuum
limit.

In Refs.~[\citen{I19a}--\citen{I22a}] we have developed the Higgs Monopole Model (HMM) approximating
the lattice artifact monopoles as fundamental pointlike particles described
by the Higgs scalar field.
Indeed, the simplest effective dynamics describing the
confinement mechanism in the pure gauge lattice U(1) theory
is the dual Abelian Higgs model of scalar monopoles~\cite{I15},\,\cite{I16}.
This model considers the following Lagrangian:
\begin{equation}
L = - \frac{1}{4g^2} F_{\mu\nu}^2(B) + \frac{1}{2} |(\partial_{\mu} -
iB_{\mu})\Phi|^2 - U(\Phi),
\lb{5y}
\end{equation}
where
\begin{equation}
U(\Phi) = \frac{1}{2}\mu^2 {|\Phi|}^2 + \frac{\lambda}{4}{|\Phi|}^4
\lb{6y}
\end{equation}
is the Higgs potential of scalar monopoles with magnetic charge $g$, and
$B_{\mu}$ is the dual gauge (photon) field interacting with the scalar
monopole field $\Phi$.  In this model $\lambda$ is the self-interaction
constant of scalar fields, and the mass parameter $\mu^2$ is negative.

Considering the renormalization group improvement
of the effective Coleman-Weinberg potential \cite{I12}, written
in Refs.~[\citen{I19a}--\citen{I22a}] for the dual sector of scalar electrodynamics in the
two-loop approximation for $\beta$-functions, we have calculated the U(1)
critical values of the magnetic fine structure constant:
\begin{equation}
{\tilde\alpha}_{\text{crit}} = g^2_{\text{crit}}/4\pi\approx 1.20
\lb{51A}
\end{equation}
and (by the Dirac relation) electric fine structure constant:
\begin{equation}
\alpha_{crit} = \pi/g^2_{\text{crit}}\approx 0.208.
\lb{51B}
\end{equation}
These values coincide with the lattice result \cite{I10a}.

Writing the following RGEs for
$\alpha_\text{i}(\mu)$ containing beta-functions for the Higgs scalar monopoles:
\begin{equation}
\frac {d(\log \alpha_\text{i}(\mu))}{dt} =
\beta(\alpha_\text{i}) - \beta^{(m)}(\widetilde \alpha_\text{i}),\quad\quad
\mbox{i=1,\,2,\,3,}
\lb{G2}
\end{equation}
we can use the one-loop approximation for $\beta(\alpha_\text{i})$ because
$\alpha_\text{i}$ are small, and the two-loop approximation for dual beta-function
$\beta^{(m)}(\widetilde \alpha_\text{i})$ by reason that $\widetilde \alpha_\text{i}$ are not very
small near the Planck scale.

It was shown in a number of investigations (see for
example~\cite{I16} and references there), that the confinement in the SU(n) lattice gauge
theory effectively comes to the same U(1) formalism. The reason is the
Abelian dominance in their monopole vacuum: monopoles of the Yang-Mills
theory are the solutions of the U(1)-subgroups, arbitrary embedded into
the SU(n) group. After a partial gauge fixing -- Abelian projection by
't Hooft~\cite{I17} -- SU(n) gauge theory is reduced to the Abelian
$U(1)^{n-1}$ theory with $n-1$ different types of Abelian monopoles.
Choosing the Abelian gauge for dual gluons, it is possible to describe
the confinement in the lattice SU(n) gauge theories by the analogous
dual Abelian Higgs model of scalar monopoles.

Using the Abelian gauge by 't Hooft and taking into account that
the direction in the Lie algebra of monopole fields are gauge
dependent, we have found in~\cite{I21a} an average over these directions and obtained
\un{the group dependence relation} between the phase transition
fine structure constants for the groups U(1) and SU(n)/Z$_\text{n}$ :
\begin{equation}
\alpha_{\text{n, crit}}^{-1} =
\frac{n}{2}\sqrt{\frac{n+1}{n-1}} \alpha_{\text{U(1), crit}}^{-1}.
\lb{25z}
\end{equation}

We have calculated this relation using only the one--loop approximation
diagrams of non-Abelian theories.

According to Eq.(\ref{25z}), we have the following relations:
\begin{equation}
\alpha_{\text{U(1), crit}}^{-1} : \alpha_{\text{2, crit}}^{-1} : \alpha_{\text{3, crit}}^{-1}
= 1 : \sqrt{3} : 3/\sqrt{2}.
\lb{26z}
\end{equation}
Near the Planck scale we are in the vicinity of the
critical points~[\citen{I19a}--\citen{I22a}].

Finally, taking into account that in the non-Abelian sectors of FRGG
we have the Abelian artifact monopoles, we obtain the following
RGEs:
\begin{equation}
\frac {d(\alpha_\text{i}^{-1}(\mu))}{dt} = \frac{b_\text{i}}{4\pi } +
\frac{N_\text{M}}{\alpha_\text{i}}\beta^{(m)}(\widetilde \alpha_{\text{U(1)}}),
\lb{G3}
\end{equation}
where $b_\text{i}$ are given by the following values:
\begin{equation}
\begin{array}{c}
b_\text{i} = (b_\text{1}, b_\text{2}, b_\text{3}) =\\
( - \frac{4}{3}N_\text{F} -\frac{1}{10}N_\text{S},\quad
\frac{22}{3}N_\text{V} - \frac{4}{3}N_\text{F} -\frac{1}{6}N_\text{S},\quad
11 N_\text{V} - \frac{4}{3}N_\text{F} ).
\end{array}
\lb{G4}
\end{equation}
The integers $N_\text{F},\,N_\text{S},\,N_\text{V},\,N_\text{M}$ are respectively the total numbers
of fermions, Higgs bosons, vector gauge fields and scalar monopoles in FRGGM
considered in our theory.

Approximating artifact monopoles by the Higgs scalar fields with a magnetic charge $g$,
we have the following Abelian monopole beta-function in the two-loop approximation~\cite{I21a}:
\be
\beta^{(m)}(\widetilde \alpha_{\text{U(1)}}) = \frac{\widetilde \alpha_{\text{U(1)}}}{12\pi }
(1 + 3 \frac{\widetilde \alpha_{\text{U(1)}}}{4\pi }).
\lb{G5}
\ee
Using the Dirac relation $\alpha \widetilde \alpha = 1/4$, we have:
\be
\beta^{(m)} =
\frac{\alpha_{\text{U(1)}}^{-1}}{48\pi }
(1 + 3 \frac{\alpha_{\text{U(1)}}^{-1}}{16\pi }),
\lb{G6}
\ee
and the group dependence relation (\ref{25z}) gives:
\be
\beta^{(m)} =
\frac{{C_\text{i}\alpha_\text{i}}^{-1}}{48\pi }
(1 + 3 \frac{{C_\text{i}\alpha_\text{i}}^{-1}}{16\pi }),
\lb{G7}
\ee
where
\begin{equation}
C_\text{i} = (C_\text{1}, C_\text{2}, C_\text{3}) =
(\frac{5}{3}, \frac{1}{\sqrt{3}}, \frac{\sqrt{2}}{3}).
\lb{G8}
\end{equation}
Finally we have the following RGEs:
\begin{equation}
\frac {d(\alpha_\text{i}^{-1}(\mu))}{dt} = \frac{b_\text{i}}{4\pi } +
N_\text{M} \frac{{C_\text{i}\alpha_\text{i}}^{-2}}{48\pi }
(1 + 3 \frac{{C_\text{i}\alpha_\text{i}}^{-1}}{16\pi }),
\lb{G9}
\end{equation}
where $b_\text{i}$ and $C_\text{i}$ are given by Eqs.(\ref{G4}) and (\ref{G8}), respectively.

In our FRGG model:
\be
\begin{array}{l}
N_{\text V}=3,\; N_{\text M}=6 \mbox{ -- for i=1,}\\
N_V = N_M =3 \;\,\mbox{\quad -- for i=2,3},
\end{array}
\lb{G10}
\ee
because we have 3 times more gauge fields ($N=3$), in comparison with usual SM
and one Higgs scalar monopole in each family.

Assuming 6 scalar Higgs bosons ($N_S=6$) breaking FRGG to SMG,
and the total number of fermions $N_\text{F}=2N_{\text{tot}}$ (usual and mirror families),
$N_{\text{tot}}=N\cdot N_{\text{gen}}=3\times 3=9$ (three SMG groups with three generations in each group),
we have obtained the evolutions of $\alpha_\text{i}^{-1}(\mu )$ near the Planck scale
by numerical calculations for $N_\text{F}=18$ and $\mu_\text{G}=10^{14}$ GeV.

Fig.2 shows the existence of the unification point. We see that in the region $\mu > \mu_\text{G}$
a lot of new fermions and a number of monopoles near the Planck scale
change the one-loop approximation behaviour of
$\alpha_\text{i}^{-1}(\mu)$ which we had in SM.
In the vicinity of the Planck scale these evolutions
begin to decrease, approaching the Planck scale $\mu = \mu_{\text{Pl}}$,
what means the suppression of the asymptotic freedom in the non-Abelian
theories.

Fig.3 demonstrates the unification
of all gauge interactions, including gravity (the intersection of
$\alpha_\text{g}^{-1}$ with $\alpha_\text{i}^{-1}$), at
\begin{equation}
\alpha_{\text{GUT}}^{-1}\approx 27 \quad
\mbox{and}                              \quad
x_{\text{GUT}}\approx 18.4.
\end{equation}

It is easy to calculate that for one family we have:
\be
\widetilde{\alpha}_{\text{GUT, one fam.}}=\frac{\ds \alpha_{\text{GUT, one fam.}}^{-1}}{\ds
4}=\frac{\ds \alpha_{\text{GUT}}^{-1}}{\ds 4\cdot 6}\approx
\frac{\ds 27}{\ds 24}\approx 1.125,
\ee
and
\be
\alpha_{\text{GUT, one fam.}}\approx 0.22,
\ee
what means that at the GUT scale electric and monopole
charges are not large and can be considered perturbatively.

Here we can expect the existence of
[SU(5)]$^3$ SUSY, or [SO(10)]$^3$ SUSY unification with superparticles of masses:
\be
M\approx 10^{18.4}\quad{\mbox{GeV}}.
\lb{G11}
\ee
The scale $\mu_{\text{GUT}}=M$, given by Eq.(\ref{G11}), can be considered
as a SUSY breaking scale.

The unification theory with [SU(5)]$^3$-symmetry was suggested first
by S.Rajpoot~\cite{I19}.

Considering the predictions of such a theory for the low-energy
physics and cosmology, maybe in future we shall be able to answer the question:
"Does the unification of [SU(5)]$^3$ SUSY or [SO(10)]$^3$ SUSY
really exist near the Planck scale?"

\section{Conclusions}

In the present paper we have shown:
\begin{itemize}
\item[1.] That the existence of monopoles in Nature
leads to the consideration of the Family Replicated Gauge Groups of symmetry
as an extension of the Standard Model in sense that using of monopoles
corresponding to the family replicated gauge fields we can bring the monopole
charge down from the unbelievably large value which it gets in the simple SM,
according to the Dirac relation.
\item[2.] If our (3+1)--dimensional space--time
is discrete and has a lattice--like structure, then the lattice artifact
monopoles play an essential role near the Planck scale if
the FRGGM works there. We have approximated these artifact
monopoles by the Higgs scalar fields.
\item[3.] The breakdown of FRGG at $\mu_\text{G}\sim 10^{14}$ GeV produces a lot of fermions
in the region $\mu_\text{G} < \mu < \mu_{\text{Pl}}$ which gives the depression  of
asymptotic freedom near the Planck scale.
\item[4.] In contrast to the AntiGUT by Froggatt--Nielsen, predicting the absence
of supersymmetry and unification up to the Planck scale,
these fermions, together with monopoles,
lead to the possible existence of unification of all interactions
(including gravity) at
$$
\mu_{\text{GUT}}=10^{18.4}\,\,GeV
$$
and
$$
\alpha_{\text{GUT}}^{-1}=27.
$$
\item[5.] The possibility of [SU(5)]$^3$ SUSY or [SO(10)]$^3$ SUSY
unifications was discussed in the present investigation.
\end{itemize}

\section{Acknowledgements}

L.V.L. and D.A.R. thank the Russian Foundation for Basic
Research (RFBR), project \#02--0217--379, for financial
support.

\newpage
\clearpage

{\Large\bf Figure captions}

\vspace{1cm}\noindent{\bf Figure 1.} The evolution of three inverse running constants
$\alpha_i^{-1}(\mu)$, where i=1,2,3 correspond to U(1), SU(2) and SU(3)
groups of the SM. The extrapolation of their experimental values
from the Electroweak scale to the Planck scale was obtained by using
the renormlization group equations with one Higgs doublet under the
assumption of a "desert". The precision of the LEP data allows to make
this extrapolation with small errors. The intersection of the inverse "gravitational finestructure
constant" $\alpha_{\makebox{g}}^{-1}(\mu)$ with $\alpha_1^{-1}(\mu)$ occurs at the
point $(x_0,\;\alpha_0^{-1})$:
$\alpha_0^{-1}\approx34.4$, and $x_0\approx18.3$, where $x=\log_{10}\mu(\text{GeV})$.\\

\noindent{\bf Figure 2.} The evolution of finestructure constants
$\alpha_{1,\, 2,\, 3}^{-1}(\mu)$ beyond the Standard model in the Family replicated gauge group model (FRGGM)
with influence of monopoles near the Planck scale.\\

\noindent{\bf Figure 3.} The evolution of $\alpha_{1,\, 2,\, 3}^{-1}(\mu)$ in
the Standard Model (SM) and beyond it. The breakdown of FRGG occurs at $\mu_G\sim 10^{14}$ GeV. It is shown
the possibility of the $[SU(5)]^3$ SUSY unification of all gauge interactions, including
gravity, at $\alpha_{GUT}^{-1}\approx 27$ and $x_{GUT}\approx 18.4$, where
$x=\log_{10}\mu$ (GeV).

\newpage
\clearpage

\noindent
\includegraphics[width=159mm, keepaspectratio=true]{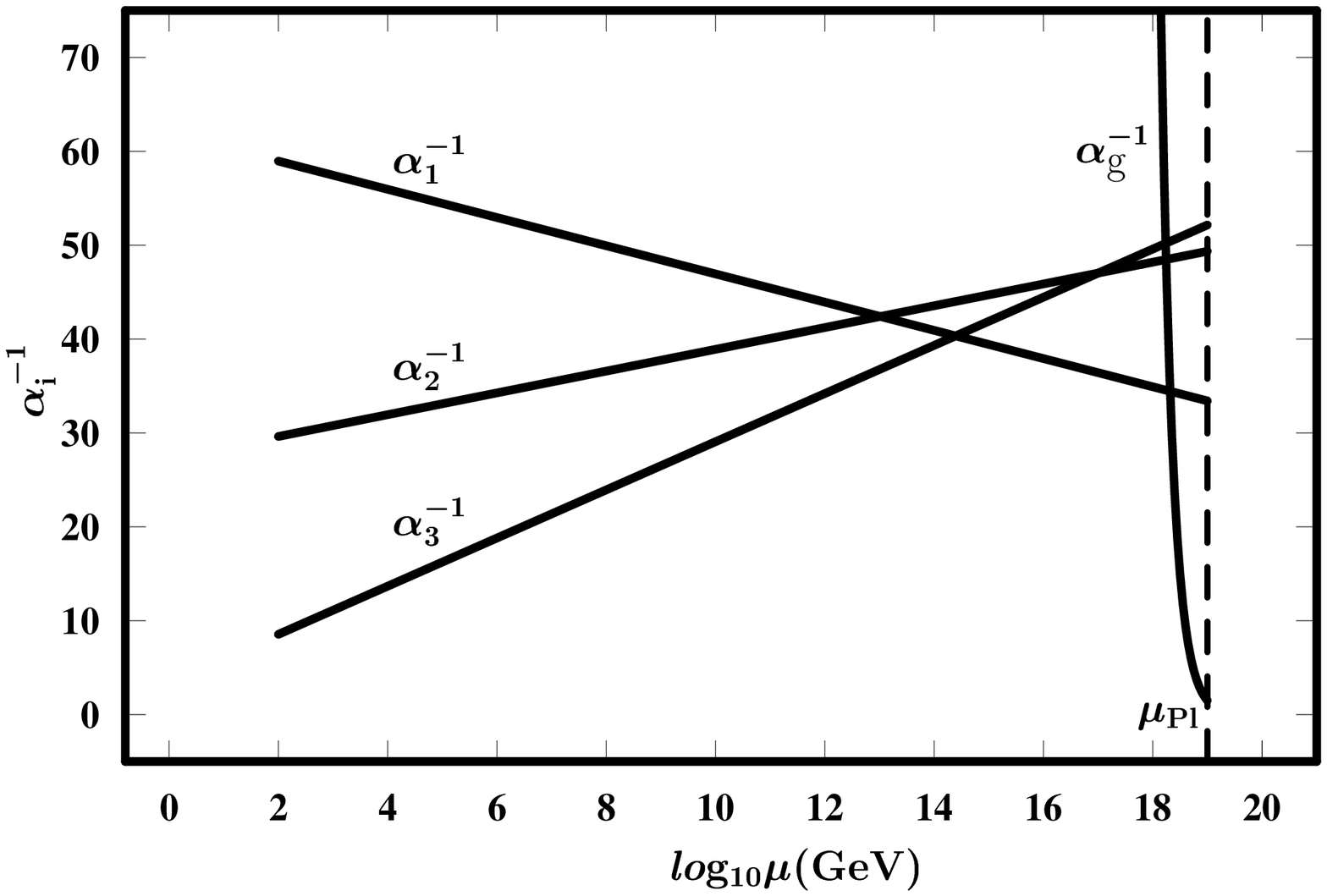}

\bc{\noindent\Large\bf Fig.1.}\ec

\newpage
\clearpage

\noindent
\includegraphics[width=159mm, keepaspectratio=true]{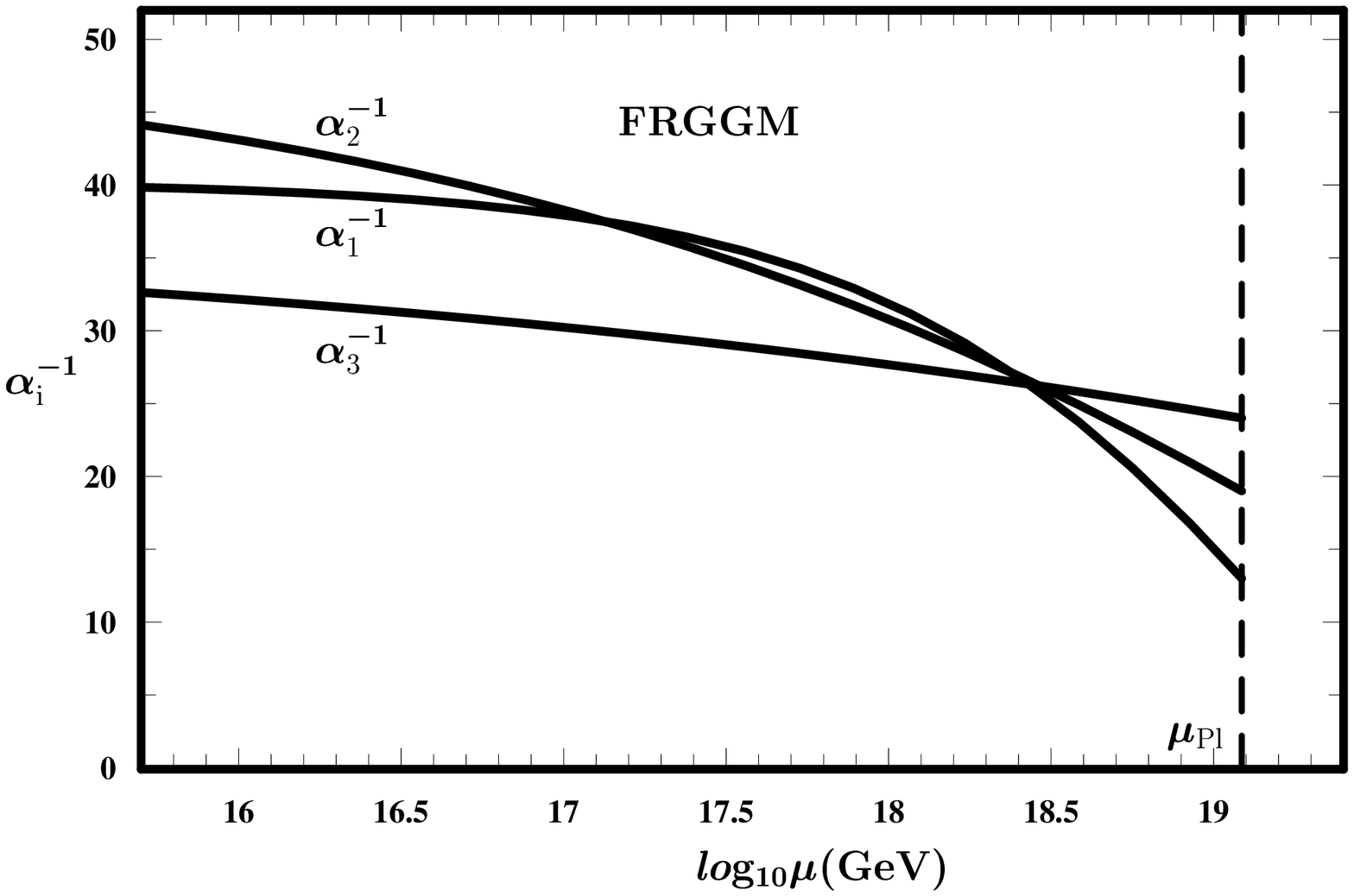}

\bc{\noindent\Large\bf Fig.2.}\ec

\newpage
\clearpage

\noindent\hspace*{-0.75pt}%
\includegraphics[width=159mm, keepaspectratio=true]{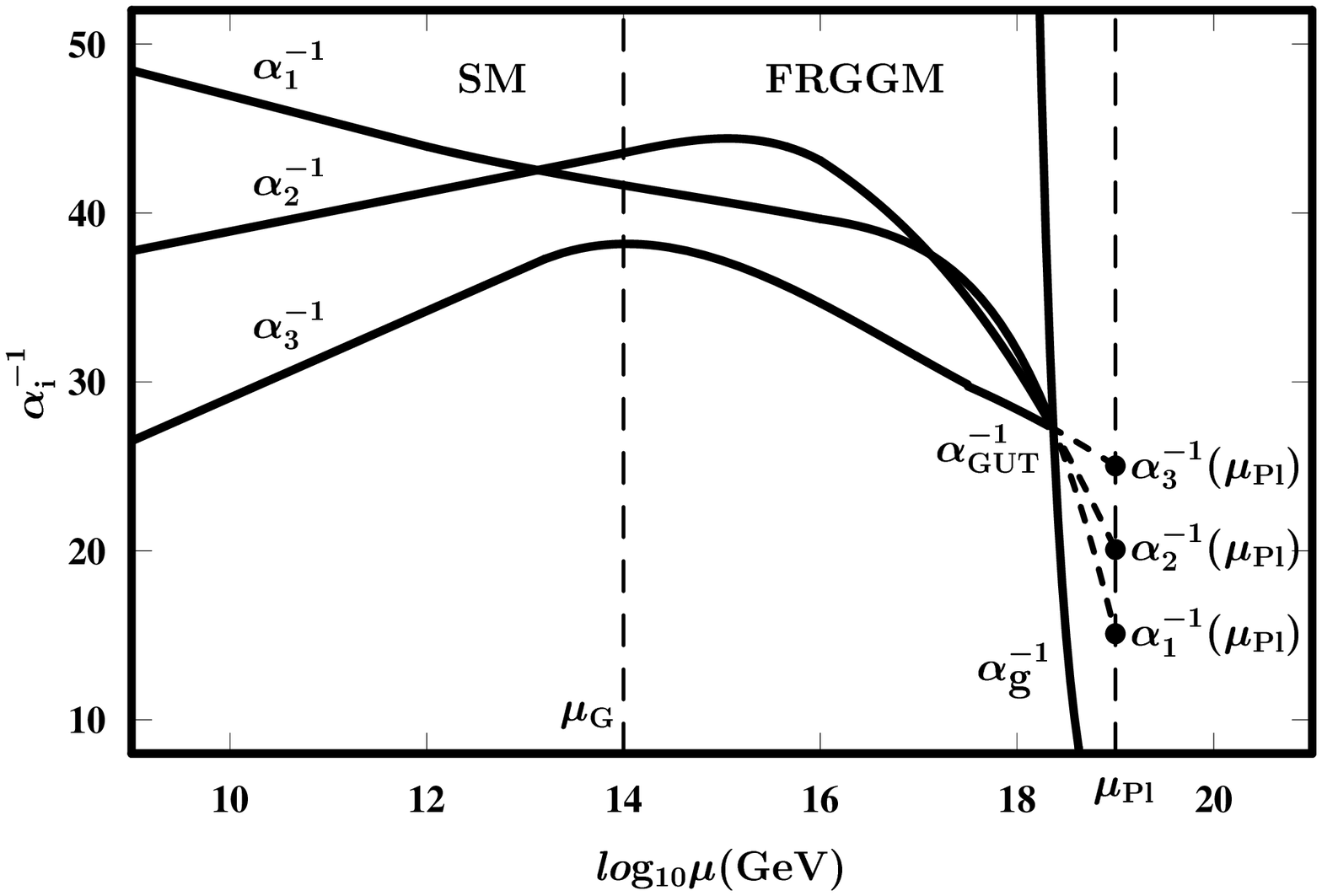}

\vspace*{0.5cm}\bc{\noindent\Large\bf Fig.3.}\ec

\end{document}